\documentclass[aps,prl,twocolumn,superscriptaddress]{revtex4}

\usepackage{graphicx}

\begin{document}

\title{An atom fiber for guiding cold neutral atoms}

\author{X. Luo}
\author{P. Kr\"uger}
\email[Electronic mail: ]{krueger@physi.uni-heidelberg.de}
\author{K. Brugger}
\author{S. Wildermuth}
\author{H. Gimpel}
\author{M. W. Klein}
\author{S. Groth}
\author{R. Folman}
\altaffiliation[Current address: ]{Department of Physics, Ben
Gurion University, Beer-Sheva 84105, Israel}
\affiliation{Physikalisches Institut, Universit\"at Heidelberg,
D-69120 Heidelberg, Germany} \homepage{www.atomchip.net}
\author{I. Bar-Joseph}
\affiliation{Department of Condensed Matter Physics, Weizmann
Institute of Science, Rehovot 76100, Israel}
\author{J. Schmiedmayer}
\affiliation{Physikalisches Institut, Universit\"at Heidelberg,
D-69120 Heidelberg, Germany} \homepage{www.atomchip.net}

\date{\today}

\begin{abstract}
We present an omnidirectional matter wave guide on an atom chip.
The rotational symmetry of the guide is maintained by a
combination of two current carrying wires and a bias field
pointing perpendicular to the chip surface. We demonstrate guiding
of thermal atoms around more than two complete turns along a
spiral shaped 25mm long curved path (curve radii down to
200$\mu$m) at various atom--surface distances (35-450$\mu$m). An
extension of the scheme for the guiding of Bose-Einstein
condensates is outlined.
\end{abstract}


\maketitle

The fast development of new tools for the precise control and
manipulation of neutral atoms makes a great variety of novel
experiments feasible. In particular, the adaption of
microfabrication techniques in atom optics laboratories has lead
to the implementation of {\em atom chips} \cite{Fol02,Rei02}. The
patterned surfaces of these devices allow trapping and guiding of
atoms with the high accuracy given by the fabrication process.
Possible applications of atom chips are abundant, fundamental
studies of degenerate quantum gases in low dimensional potentials
and mesoscopic physics in small atomic ensembles being just two
prominent examples. In the quest for implementations of quantum
information processing (QIP) with neutral atoms, atom chips are
especially promising candidates.

The versatility of atom chips has been shown in a number of
experiments. After the demonstration of simple magnetic trapping
and guiding potentials \cite{Rei99,Fol00}, the production of
Bose-Einstein condensates (BEC)\cite{Ott01,Hae01b,Lea02,Sch03} and
the integration of electrostatic fields \cite{Kru03} on the chips
have been important milestones on the way to a fully functional
toolbox for the control of atomic matter waves. Issues currently
under investigation include the integration of light elements for
enhanced detection efficiency of (single) atoms \cite{Hor03} and
the coherence properties of atoms in the chip potentials
\cite{Hen03}.

\begin{figure}
    \includegraphics[width = \columnwidth]{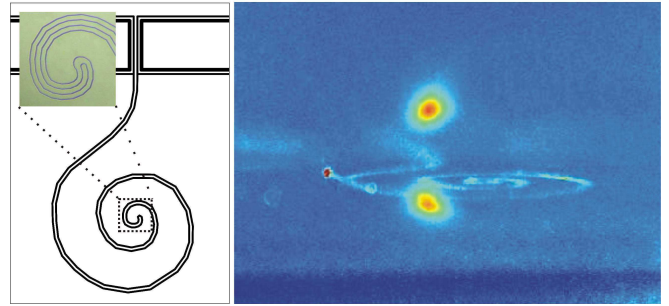}
    \caption{(left) Chip design used in the experiment. The black
    lines indicate current carrying gold wires, the white areas
    are grounded parts of the chip surface. The insert shows a microscope
    image of a detail of the spiral shaped wire guide. The 10$\mu$m wide
    grooves from which the gold has been removed to define the
    wires show as dark lines. (right) Fluorescence of a
    magnetically trapped cloud and its reflection from the chip
    surface just before the guide is loaded. The guiding wires
    are visible through scattered imaging light.
    \label{fig:design}}
\end{figure}

Here, we report on the implementation and experimental test of a
key element for the controlled manipulation of matter waves on the
atom chip: an omnidirectional `atom fiber', i.e. an atomic wave
guide based on a potential that is independent of the guiding
direction (Fig.\ \ref{fig:design}). The use of such an element is
inevitable when atoms are to be transported to and stored in
individual trapping sites on the two dimensional surface of the
chip. This will be of particular significance for implementations
of quantum registers for QIP based on a 2-dimensional array of
microtraps. Furthermore, the directional symmetry of this type of
guide is crucial for guided matter wave interferometers that rely
on spatially symmetric beam splitters, i.e (wide angle)
Mach-Zehnder and Sagnac interferometers.

\begin{figure}
    \includegraphics[width = \columnwidth]{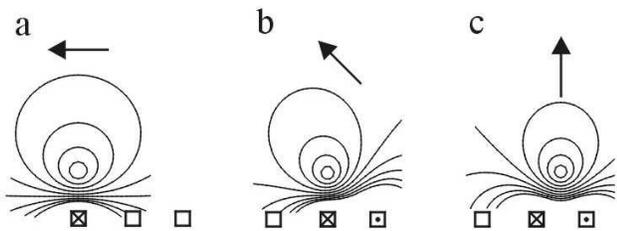}
    \caption{Potential configurations during the transfer of atoms
    from a single-wire guide with horizontal bias field (a) to a
    two-wire guide with vertical bias field (c). In each
    configuration, an arrow points in the direction of the bias field and
    three squares represent the three wires. The
    current flow is indicated by the symbols in the squares. A dot
    (cross) corresponds to a current flow out of (into) the shown
    plane, a blank square corresponds to zero current. In the
    intermediate stage (b), the currents run already exclusively through the
    two wires carrying counter-propagating currents while the bias
    field has been rotated by $45^\circ$ with respect to the wire plane.
    \label{fig:loading}}
\end{figure}

The simplest form of a magnetic wire guide is the {\em side guide}
in which atoms in weak field seeking states ($U_\mathrm{mag}
=-\mbox{\boldmath$\mu$} \cdot {\mathbf B} > 0$) are trapped in a
potential tube along a line parallel to a straight current
carrying wire (Fig.\ \ref{fig:loading}a). To achieve this
configuration, the field of the wire is superimposed with a
homogenous external bias field. Atoms can only move along the
third unconfined direction as long as the bias field points in the
direction perpendicular to the guiding direction. This implies
that the side guide is only a single-directional guide
\footnote{Small angular deviations from the straight guiding path
can be tolerated for thermal atoms as long as the potential
barriers along the guide are small compared to the atom cloud
temperature.}. True multi-directional guiding is only possible for
guides based on at least two current carrying wires. One possible
two-wire guide is based on two parallel wires carrying {\em
co}-propagating currents. In this case, no additional bias field
is needed and the guide can be shaped simply by bending the wires.
Wire guides with two co-propagating currents have been realized
and it has been shown that atoms can be deflected by small angles
\cite{Mue99} and even be stored in a complete loop where a cloud
has been observed to move around a number of full turns
\cite{Sau01}. A drawback, however, lies in the fact that the
rotational symmetry of the potential can only be maintained when
no external bias field is added. This implies that the potential
minimum will always be located in the center between the two wires
which limits the flexibility severely.

An improved type of two-wire guide is based on parallel {\em
counter}-propagating currents (Fig.\ \ref{fig:loading}c). Here,
the field of the wires can be compensated by a field pointing {\em
perpendicular} to the wire plane which avoids the usual symmetry
breaking associated with the addition of an external homogenous
field. Together with the currents, the strength of the bias field
determines the height of the guide {\em above} the wire plane and
the potential parameters \cite{Thy99a}. In a first experiment, a
free falling atomic cloud has been observed to be guided by a
straight two-wire guide with counter-propagating currents
\cite{Dek00}.

In our experiment, we set out to demonstrate deterministic loading
and actual guiding of atoms confined in a bent two-wire guide with
counter-propagating currents. For this purpose, we designed a
spiral shaped two-wire guide (Fig.\ \ref{fig:design}). The two
wires (width $\times$ height $=45\times 5\mu$m$^2$, center to
center spacing $2d=115\mu$m) are connected at the inner end of the
spiral. This automatically leads to a counter-propagating current
flow. The spiral shape was chosen in order to demonstrate the full
flexibility of the guide by incorporating more than two full
rotations with curve radii ranging from 200$\mu$m to 3mm along the
25mm long guiding path. The U-shaped wires (cross section
$200\times 5\mu$m$^2$) on either side of the straight beginning of
the guide are used to form three dimensional traps \cite{Fol02}.

The starting point of our atom chip experiments is a reflection
magnetooptical trap (MOT) \cite{Rei99} that contains a cloud of
typically $10^8$ cold $^7$Li atoms located a few millimeters above
the chip surface. The magnetic quadrupole field for the MOT is
initially provided by external coils and after the loading
replaced by the field of a U-shaped wire ($\emptyset=1$mm) mounted
directly underneath the chip. This allows to bring the atoms
closer to the surface and to transfer them to a purely magnetic
trap formed by the same magnetic field. In the next step, the
cloud is loaded to the magnetic potentials produced by wires on
the chip. The details of this procedure are given in
\cite{Fol00,Cas00}.
\begin{figure}
    \includegraphics[width = \columnwidth]{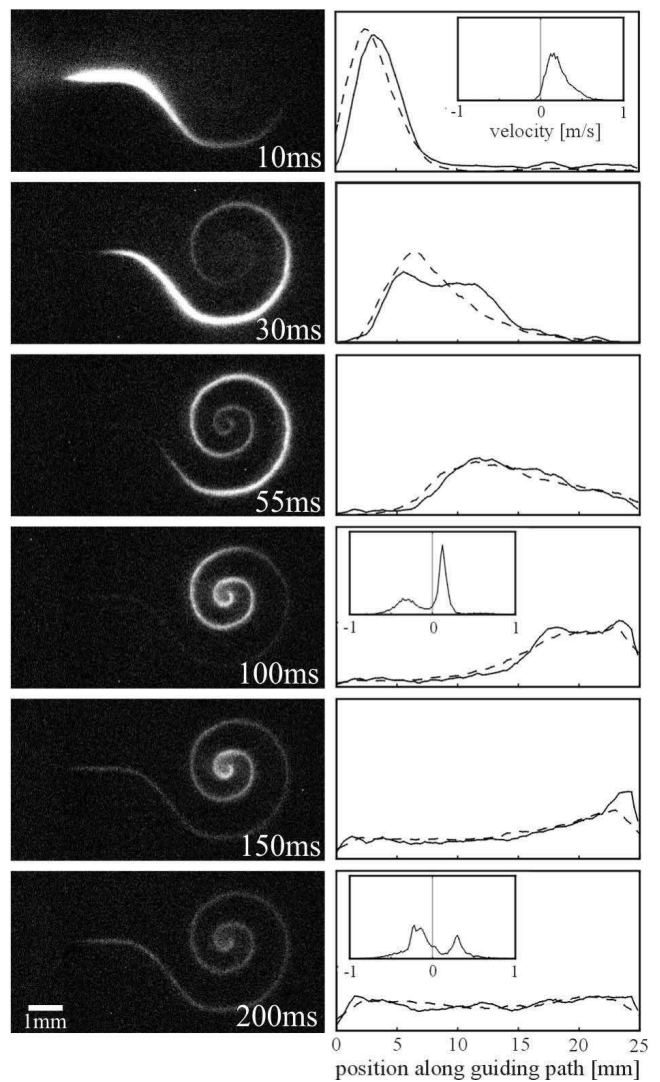}
    \caption{Time sequence of atoms released from the reservoir trap
    into the spiral shaped guide. (left) Fluorescence images.
    Atoms that have reached the end of the guide (center of the
    spiral) are reflected from a potential barrier and propagate in
    the backward direction. (right) One dimensional density
    distributions along the path of the spiral, extracted from the
    experimental data and Monte-Carlo (MC) simulations (solid and dashed
    curves, respectively). In the inserts, the corresponding velocity
    distributions obtained by the same MC calculations
    are depicted. In these plots, a clear signature of the
    reflection is visible, the part of the cloud propagating
    backwards is clearly separated from that propagating in the
    forward direction.
    \label{fig:movie}}
\end{figure}
In order to transfer the atoms from a U-wire trap, i.e. a trap
based on a single wire and a horizontal bias field, to the spiral
shaped two-wire guide, we ramp down the current of the single wire
while ramping up the counter-propagating currents in the two
parallel wires of the guide. During this first step, the bias
field is only partially rotated so that the bending of the wires
still provides an endcap of the potential, thus confining the
atoms in three dimensions. As depicted schematically in Fig.\
\ref{fig:loading}b, this intermediate configuration is reminiscent
of the simple side guide (Fig.\ \ref{fig:loading}a) with only a
slight perturbation by the current in the extra wire. In the final
step, the rotation of the bias field is completed (Fig.\
\ref{fig:loading}c), and the atoms can expand freely along the
spiral shaped path of the guide.

Fig.\ \ref{fig:movie} shows a time sequence of the fluorescence
signal of atoms in the guide \footnote{The images are taken by
exposing the atoms to a flash (100$\mu$s) of near resonant laser
light. In order to avoid any disturbing reflections, the light
enters the chamber from two directions parallel to the chip
surface.}. Guiding of atoms was possible over a wide range of
parameters. By varying the bias field strength $B$ from 1G to 50G
at a constant current of 1A through both (connected) wires, the
height of the potential tube above the surface was scanned from
450$\mu$m to 35$\mu$m and corresponding gradients of 40G/cm to
8kG/cm for atoms in the $|F=2,m_F=2\rangle$ state. The images and
density profiles in Fig.\ \ref{fig:movie} show that the atom cloud
not only expands according to its temperature \footnote{The clouds
exhibit an anisotropic temperature profile (450$\mu$K in the
transverse, 50$\mu$K in the longitudinal direction) due to a
transverse compression during the loading without
rethermalization.} but also moves as a whole along the guide. This
center of mass motion is induced by a longitudinal field gradient
produced by the current in the two leads from the beginning of the
spiral wires to the connecting pads on the edge of the chip
\footnote{By running a parallel current through another wire on
the chip, we could even enhance this `pushing' effect.}.

For a quantitative understanding of the density profiles, we
performed Monte-Carlo (MC) simulations of classical trajectories
of particles in the guide. The results are depicted in Fig.\
\ref{fig:movie} and show good agreement with the experiment. In
particular, the reflections that occur when atoms reach the inner
end of the guide, are reproduced well. The effect of the
reflection becomes most apparent in the plots of the velocity
distributions extracted from the MC calculations (inserts). The
velocity classes for forward and backward motion in the guide are
clearly separated.

\begin{figure}
    \includegraphics[width = \columnwidth]{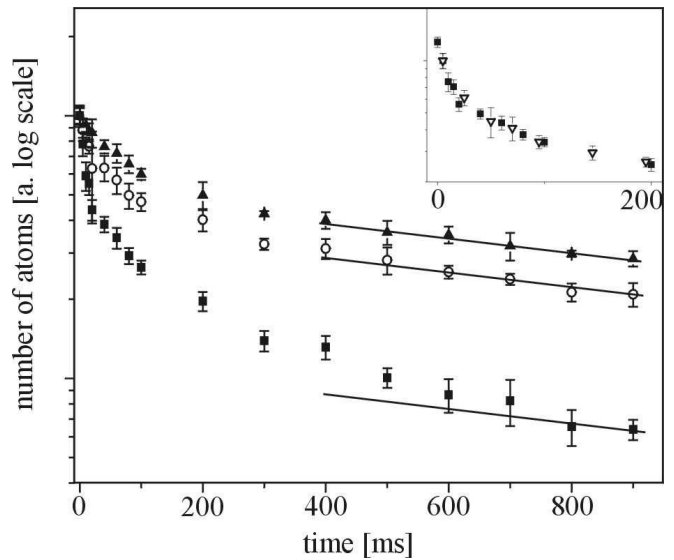}
    \caption{Lifetime graphs of three dimensional traps based on
    two counter-propagating currents in combination with a
    vertical bias field. Two different loss mechanisms can be
    distinguished in the time domain: An initial overexponentially
    fast process plays a role during the first $\sim 100$ms;
    afterwards a slow exponential loss (time constant 1.6s, solid lines) stemming
    from collisions with the background gas dominates. Both the
    rate and amount of the fast loss depend on the depth of the
    trapping potential while the vacuum limited loss does not. The
    data were taken for traps with depths of $k_B \times$
    1250, 950, 500$\mu$K (triangles, circles, squares, respectively).
    The insert shows a comparison between the lifetime of a three
    dimensional trap (squares) and the guide (triangles)
    of approximately equal potential depth of $k_B \times$ 500$\mu$K.
    \label{fig:lifetime}}
\end{figure}

For a characterization of the guides, lifetimes of confined atom
clouds are of great significance. In our single vacuum chamber
apparatus, the rest gas pressure is typically of the order of
$\sim 10^{-9}$mbar, corresponding to background gas collision
limited lifetimes of $\sim1$s as was confirmed in a conventional
single Z-wire trap \cite{Fol02}.

Direct lifetime measurements in the guide were only carried out
for guiding times of up to $\sim 200$ms \footnote{Ohmic heating in
the long guiding wires did not allow longer guiding times without
risking damage to the wires.}. Indirect lifetime measurements were
performed in three dimensionally confining potentials based on
counter-propagating currents through two parallel wires. For this
purpose we used the two U-shaped wires (Fig.\ \ref{fig:design})
where the confinement in the longitudinal direction is provided by
the wire leads. In these traps, the lifetime measured in the
conventional Z-trap could be reproduced for the long trapping time
regime ($\tau>$300ms). For shorter times, the shape of the
potential leads to a faster additional loss of the hottest atoms
on a timescale of $\sim 100$ms. This behavior was again confirmed
by MC-calculations and will not pose a problem for colder atom
samples. The measured results for different potential depths are
depicted in Fig.\ \ref{fig:lifetime}. The comparison of the data
obtained for the guide and for the three-dimensional trap (insert)
shows complete agreement on the short timescale indicating the
validity of the indirect measurements also for longer times.

In future applications of vertical bias field guides, it will be
crucial to be able to guide much colder atoms (ideally BEC) than
the ones used in our demonstration and characterization. In this
case, a loss mechanism that does not play a role in the
experiments presented here will be dominating: Majorana spin flip
transitions will remove atoms from the guide in the vicinity of
the quadrupole minimum of the potential where the field vanishes.
The usual remedy employed in single-directional guides is the
addition of a small field component pointing in the guiding
direction (Ioffe-Pritchard (IP) field). For a multi-directional
guide this would mean that a locally varying field would have to
be used or, alternatively, the IP field could be rotated as a
(small) cloud moves along the guide.

\begin{figure}
    \includegraphics[width = \columnwidth]{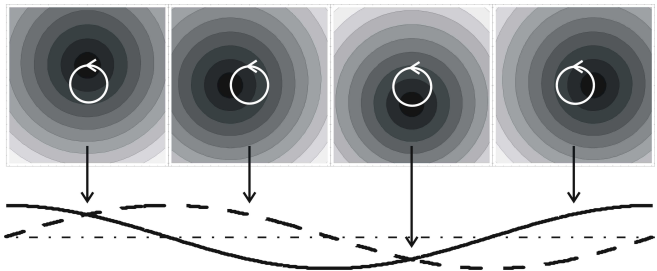}
    \caption{(top) Time sequence over one oscillation period of a time
    orbiting guiding potential. Darker shading corresponds to
    lower potential. (bottom) Each of the counter-propagating
    currents (solid and dashed curves)
    in two parallel wires is sinusoidally modulated around the
    steady current $I_0$ (dashed-dotted line). A relative
    phase difference of $\Delta\phi=\pi/2$ results in a quadrupole field zero
    circling around the minimum of the static situation (white arrows). With a
    proper choice of the modulation frequency, cold atoms are
    trapped in a time averaged potential. While the position of
    the potential minimum remains unchanged with respect to the
    static case, the atoms never encounter a
    magnetic field zero and thus do not undergo Majorana spin flips.
    \label{fig:top}}
\end{figure}

Here, we suggest a solution that we believe to be more flexible
and easier to implement experimentally. Sinusoidally modulating
the currents  according to $I(t)=I_0+I_\mathrm{mod}
\sin(\omega_\mathrm{mod} t+\phi)$ in the two (now separated)
guiding wires with a relative phase difference of
$\Delta\phi=\pi/2$ leads to a (nearly) circular motion of the
quadrupole minimum of the guide (Fig.\ \ref{fig:top}). As long as
the modulation frequency $\omega_\mathrm{mod}$ is slow with
respect to the Larmor frequency $\omega_\mathrm{Lar}$ but fast
with respect to the atomic oscillation frequency
$\omega_\mathrm{trap}$, the atoms can be described as moving in a
time averaged (orbiting) potential (TOP). Such TOP are routinely
used for the production of Bose-Einstein condensates \cite{Pet95}.
If the height $h$ of the guide over the wire plane is equal to the
half distance $d$ between the wires, the parameters of the
averaged potential are $\omega_\mathrm{Lar}=\frac{g_F \mu_B B
I_\mathrm{mod}}{\sqrt{2} \hbar I_0}$ and
$\omega_\mathrm{trap}=\frac{2 \pi}{\mu_0} \sqrt{\frac{g_F m_F
\mu_B}{\sqrt{2} M}\frac{B^3}{I_0 I_{\mathrm{mod}}}}$ with the
vacuum permeability $\mu_0$, the Bohr magneton $\mu_B$, the
Land\'{e} factor $g_F$, and the atomic mass $M$. The depth of the
potential is given by the radius of the circle of vanishing field
(`circle of death') $r_0=d I_\mathrm{mod}/\sqrt{2} I_0$. The
potential tuning range is large, a TOP two wire ($d=20\mu$m) atom
fiber for a BEC with $\omega_\mathrm{Lar}=2\pi\times 500$kHz,
$\omega_\mathrm{trap}=2\pi\times 5$kHz and $r_0=1.4\mu$m could,
for example, be obtained with $I_0=100$mA, $I_\mathrm{mod}=10$mA,
and $B=10$G ($^{87}$Rb atoms in the in the $|F=2,m_F=2\rangle$
state).

To conclude, we have demonstrated the controlled loading and
guiding of atoms in a truly omnidirectional guide that exhibits
complete rotational symmetry. The flexibility of the guide was
proven by operating it over a wide parameter range. Monte-Carlo
simulations reproduce the measured time dependent atomic density
profiles well. We have discussed a scheme involving time dependent
currents to modify the guide in order to lift the current
restriction to thermal atoms. Future applications range from the
loading of two-dimensional trap arrays to the realization of
circular of wide angle matter wave interferometers \cite{And02}.

\begin{acknowledgments}
This work was supported by the European Union, contract numbers
IST-2001-38863 (ACQP), HPRI-CT-1999-00114 (LSF) and HPRN-CT
2002-00304 (FASTNet) and the Deutsche Forschungsgemeinschaft,
Schwerpunktprogramm `Quanteninformationsverarbeitung'.
\end{acknowledgments}

\end{document}